\documentclass{elsart5p}
\usepackage{graphics}
\usepackage{graphicx}
\usepackage{epsfig}
\usepackage{amssymb}
\usepackage{amsmath}
\usepackage{color}

\newcommand{\dd}{\mbox{$\textrm{d}$}}

\begin{document}
\begin{frontmatter}

\title{Comparison of the $pp\to\pi^+ pn$ and $pp\to \pi^+d$ production rates}
\author[inst1]{G.~F\"{a}ldt}\ead{goran.faldt@physics.uu.se}
and
\author[inst2]{C.~Wilkin\corauthref{cor1}}\ead{c.wilkin@ucl.ac.uk}
\corauth[cor1]{Corresponding author at: Physics and Astronomy Department, UCL, Gower Street, London, WC1E 6BT, UK}
\address[inst1]{
Department of Physics and Astronomy, Uppsala University, Box 516, 751 20 Uppsala, Sweden}
\address[inst2]{
Physics and Astronomy Department, UCL, Gower Street, London, WC1E 6BT, UK }
\date{\today}
\begin{abstract}Fully constrained bubble chamber data on the $pp\to \pi^+pn$ and
$pp\to \pi^+d$ reactions are used to investigate the ratio of the counting
rates for the two processes at low $pn$ excitation energies. Whereas the
ratio is in tolerable agreement with that found in a high resolution
spectrometer experiment, the angular distribution in the final $pn$ rest
frame shows that the deviation from the predictions of final state
interaction theory must originate primarily from higher partial waves in the
$pn$ system. These considerations might also be significant for the
determination of the $S$-wave $\Lambda p$ scattering length from data on the
$pp\to K^+\Lambda p$ reaction.
\end{abstract}
\end{frontmatter}

The COSY-GEM collaboration measured the differential cross section for the
production of positive pions in proton-proton collisions at a beam kinetic
energy of $T_p=951$~MeV, detecting the $\pi^+$ at $\theta_{\pi}=0^{\circ}$ in
the spectrograph Big Karl~\cite{ABD2005}. The high missing-mass resolution
achieved here ($\approx 100$~keV/$c^2$) allowed a very clean separation of
the $pp\to \pi^+d$ and $pp\to\pi^+ pn$ channels and also showed that the
production of singlet $pn$ pairs was negligible at this energy.

The final state interaction theorem relates the normalisations of the wave
functions for $S$-wave bound and scattering states~\cite{FAL1997}. This has
been exploited to predict the double-differential centre-of-mass (cm) cross
section for the $S$-wave spin-triplet component in $pp\to\pi^+pn$ in terms of
the cross section for $pp\to\pi^+d$~\cite{BOU1996}:
\begin{eqnarray}\nonumber
\frac{\dd^2\sigma}{\dd\Omega\,\dd{x}}
(pp\to\pi^+\left\{pn\right\}_t) &=&\\
&&\hspace{-2cm}\mathcal{F}\,\frac{p(x)}{p(-1)}
\frac{\sqrt{x}}{2\pi(x+1)}\,\frac{\dd\sigma}{\dd\Omega}(pp\to
\pi^+d)\:.\label{equ:d_pn}
\end{eqnarray}
Here $x$ denotes the excitation energy $Q$ in the $np$ system in units of the
deuteron binding energy $B_t$, $x=Q/B_t$, and $p(x)$ and $p(-1)$ are the pion
cm momenta for the $pn$ continuum or deuteron respectively. At the deuteron
pole the normalisation $\mathcal{F}=1$ but it was argued~\cite{FAL1997} that
deviations from this should be small at low $x$ if the pion production
operator is of short range and the tensor force linking the $S$ and $D$
states in the deuteron could be neglected. However, although the shape of the
COSY-GEM data~\cite{ABD2005} was reasonably well described by
Eq.~(\ref{equ:d_pn}) up to an excitation energy of $Q\approx 20$~MeV,
reproducing the absolute magnitude required $\mathcal{F}=2.2\pm 0.1$.

In view of the large discrepancy with the prediction of the final state
interaction theorem ($\mathcal{F}=1$), the COSY-GEM experiment was repeated
at 400 and 600~MeV~\cite{BUD2009} which, combined with the results of earlier
work carried out at TRIUMF~\cite{PLE1999}, presented a consistent picture.
The normalisation factor $\mathcal{F}$ was found to increase steadily from
below one at 400~MeV to well above unity at 951~MeV. It was suggested that
the deviation at the highest energy could arise from the long-range part of
the pion production operator associated with the on-shell intermediate
pions~\cite{BUD2009}. Such contributions could change the $pn$ $S$-wave cross
section for the $pp\to\pi^+pn$ reaction or excite higher partial waves in the
final $pn$ system. In a missing-mass experiment, as carried out by the GEM
collaboration~\cite{ABD2005,BUD2009}, it is not possible to investigate these
suggestions any further.

Measurements of various channels arising from proton-proton collisions at
three energies in the 900 to 1000~MeV range were undertaken using the 35~cm
hydrogen bubble chamber of the Petersburg Nuclear Physics Institute
(PNPI)~\cite{ERM2014,ERM2011,ERM2017}. Although the statistics in the low $Q$
region are much poorer than those of the COSY-GEM
experiment~\cite{ABD2005,BUD2009}, and the resolution is far inferior, the
acceptance approaches 100\% and so the predictions of Eq.~(\ref{equ:d_pn})
can be integrated over the full solid angle. Under the conditions of the PNPI
data, the deviation of $p(x)/p(-1)$ from unity is negligible at low $x$ so
that the numbers $N$ of bubble chamber events should be linked by
\begin{eqnarray}\nonumber
N_{x<x_0}(pp\to\pi^+\left\{pn\right\}_t) &=&
\frac{\mathcal{F}}{2\pi}\,N(pp\to\pi^+d)\int_0^{x_0}
\frac{\sqrt{x}\,\dd x}{(x+1)}\\
&&\hspace{-3cm}=\frac{\mathcal{F}}{\pi}\,N(pp\to\pi^+d)\left(\sqrt{x_0}-\arctan(\sqrt{x_0})\right).
\label{equ:d_pn1}
\end{eqnarray}

In order to compare directly the PNPI data with the COSY-GEM result, the
bubble chamber $pp\to\pi^+pn$ events were selected as having a maximum $pn$
excitation energy of 20~MeV ($x_0 \approx 9$). The numbers of events
fulfilling this criterion, as well as the total number of $pp\to\pi^+d$ events,
are given in Table~\ref{Results}. The values of $\mathcal{F}$ deduced at the
three energies are also shown, as is their average.

\begin{table}[hbt]
\centering \caption{The numbers of events measured in the $pp\to\pi^+d$ and
$pp\to \pi^+pn$ (with $pn$ excitation energy below 20~MeV) reactions in the
PNPI bubble chamber experiment at 900.2~MeV~\cite{ERM2014},
940.7~MeV~\cite{ERM2011}, and 988.6~MeV~\cite{ERM2017}. The values derived
for the normalisation factor $\mathcal{F}$ and its average over the three
energies are also listed. Only statistical errors are quoted. The results
shown take into account the small numbers of ambiguous events.}
\label{Results}
\begin{tabular}{|c|c|c|c|} \hline
$T_p$(MeV)&$N(pp\to\pi^+pn)$&$N(pp\to\pi^+d)$&$\mathcal{F}$\\
\hline
900.2&188&136&$2.6\pm0.3$\\
940.7&101&\phantom{1}77&$2.4\pm0.3$\\
988.6&\phantom{1}97&\phantom{1}57&$3.1\pm0.5$\\
\hline
Summed&386&270&$2.6\pm0.2$\\
\hline
\end{tabular}
\end{table}

The weighted average of the PNPI data, $\mathcal{F}=2.6\pm0.2$, looks higher
than the COSY-GEM value of $2.2\pm0.1$ but the errors quoted are only
statistical uncertainties. Furthermore, it is possible that the fraction of
higher $pn$ waves could vary with pion angle. However, there might also be an
effect due to the poorer resolution in the $pn$ excitation energy in the
bubble chamber data, which allows some higher $x$ data to distort a little
the result.

\begin{figure}[htb]
\begin{center}
\includegraphics[width=1.0\columnwidth]{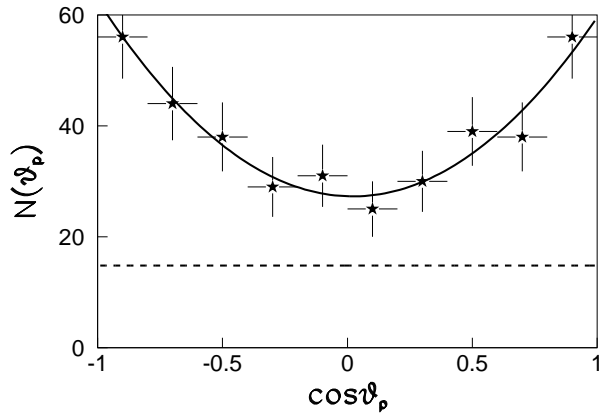}
\caption{\label{fig:Ang} Numbers of $pp\to\pi^+pn$ events with statistical
errors measured in the PNPI bubble chamber experiment, summed over three
incident beam energies~\cite{ERM2014,ERM2011,ERM2017}. Events, which are
shown in $0.2$ bins in $\cos\theta_p$, are only retained where the excitation
energy in the $pn$ system is below 20~MeV. The solid curve shows the best
quadratic fit of Eq.~(\ref{quadratic}) whereas the dashed line is the
isotropic $S$-wave prediction that follows from the final state interaction
theorem~\cite{FAL1997,BOU1996}.}
\end{center}
\end{figure}

The big advantage of the fully constrained PNPI $pp\to \pi^+pn$ data is that
they allow one to investigate the angular distributions in the recoiling $pn$
system\footnote{The measured four-vectors in the PNPI $pp\to\pi^+pn$
experiment at the three different beam energies are listed on the WEB site of
the Bonn-Gatchina Partial Wave Analysis group: pwa.hiskp.uni-bonn.de.}.
Figure~\ref{fig:Ang} shows the distribution in the angle of the final proton
with respect to the original beam direction in the $pn$ rest frame. The clear
deviation from isotropy is unambiguous evidence for the production of higher
partial waves in the final $pn$ system.

The data in Fig.~\ref{fig:Ang} are well described by the form
\begin{equation}
\label{quadratic}
N(pp\to\pi^+pn) = a_0 +a_1\cos\theta_p +a_2\cos^2\theta_p\,,
\end{equation}
where $a_0=27.3\pm2.7$, $a_1=-1.5\pm3.7$, and $a_2=33.8\pm7.1$. The
corresponding value of $\chi^2/\textrm{NDF} =0.27$ is fortuitously low. It is
important to note that, as expected, the odd term $a_1$ is consistent with
zero but fixing it to vanish does not change significantly the values of
$a_0$ and $a_2$. However, it is not possible to identify whether the large
quadratic term arises from the square of a $pn$ $P$-wave or from an $S-D$
interference, which could be influenced by the $pn$ tensor force. Departures
from isotropy in Fig.~\ref{fig:Ang} are clear evidence for the excitation of
higher partial waves but the converse is not true because it is possible to
generate a mixture of higher partial waves that leads to an isotropic
distribution. Nevertheless, it is likely that the deviations from the final
state interaction theorem of Refs.~\cite{FAL1997,BOU1996}, shown by the
dashed line in Fig.~\ref{fig:Ang}, probably come from higher partial waves
rather than modifications of the $S$-wave intensity.

Apart from complications arising from the $pn$ tensor force, it is to be
expected that Eq.~(\ref{equ:d_pn}) should be a good representation of the
$pp\to\pi^+pn$ data at very small values of $x$. Although this is a valid
approximation at low incident beam energies, the COSY-GEM experiment shows
that there are significant deviations at 951~MeV~\cite{ABD2005}. By using the
fully reconstructed bubble chamber events~\cite{ERM2014,ERM2011,ERM2017}, we
have confirmed the magnitude of the deviation. However, we have also shown
from the angular distribution of Fig.~\ref{fig:Ang} that, with a cut-off at
$Q=20$~MeV, there are significant contributions from higher partial waves in
the $pn$ system that are not apparent in a missing-mass experiment. This may
be due to the anomalously long range of the pion production operator at high
energies which was not considered in the application of the final state
interaction theorem~\cite{FAL1997}.

The arguments presented here may have wider significance than the specific
reaction being studied. By detecting just the $K^+$ meson in the Big Karl
spectrograph, the COSY-HIRES group measured the inclusive cross section for
the $pp\to K^+X$ reaction. Below the threshold for $\Sigma$ production
$X=\Lambda p$ and the hope was that an analysis of the data would allow a
determination of the spin-average $\Lambda p$ $S$-wave scattering
length~\cite{BUD2010}. However, in such a single-arm experiment, there can be
no confirmation that the $\Lambda p$ system remains in the $S$-wave at finite
values of $Q$.

Conditions are much more favourable in the COSY-TOF
experiment~\cite{HAU2017,HAU2014} where, apart from some loss of acceptance
near the beam direction, the final particles in the $pp\to K^+\Lambda p$
reaction can be detected. The global $\Lambda p$ angular distributions
constructed for beam momenta of 2.7~GeV/$c$~\cite{HAU2014} and
2.95~GeV/$c$~\cite{JOW2016} both show strong signals arising from higher
partial waves in the $\Lambda p$ system but it is not clear from these plots
if there are also effects for $\Lambda p$ excitation energies below 40~MeV.
This clearly has to be checked when attempting to extract the $S$-wave
$\Lambda p$ scattering length from the analysis of such
experiments~\cite{BUD2010,HAU2017}. The determination of the position of the
$\Lambda p$ virtual state pole~\cite{FAL2017} is far less affected by these
considerations because this is sensitive to the behaviour at very small
values of $Q$, where the $S$-wave assumption is on much firmer grounds.

\newpage
Correspondence with F.~Hauenstein and J.~Ritman has been very instructive.
The cooperation of A.~Sarantsev and V.~Sarantsev has been most helpful.


\begin{thebibliography}{99}
%
\bibitem{ABD2005} M.~Abdel-Bary et al., Phys.\ Lett.\ B 610 (2005) 31.
%
\bibitem{FAL1997} G.~F\"{a}ldt and C.~Wilkin, Physica Scripta 56 (1997) 566.
%
\bibitem{BOU1996} A.~Boudard, G.~F\"{a}ldt, and C.~Wilkin: Phys.\ Lett.\ B 389 (1996) 440.
%
\bibitem{BUD2009} A.~Budzanowski et al., Phys.\ Rev.\ C 79 (2009) 061001(R).
%
\bibitem{PLE1999} R.~G.~Pleydon et al., Phys.\ Rev.\ C 59 (1999) 3208.
%
\bibitem{ERM2014} K.~N.~Ermakov et al., Eur.\ Phys.\ J.\ A 50 (2014) 98.
%
\bibitem{ERM2011} K.~N.~Ermakov et al., Eur.\ Phys.\ J.\ A 47 (2011) 98.
%
\bibitem{ERM2017} K.~N.~Ermakov et al., arXiv:1702.05385 [nucl-ex]  (2017).
%
\bibitem{BUD2010} A.~Budzanowski et al., Phys.\ Lett.\ B 687 (2010) 31.
%
\bibitem{HAU2017} F.~Hauenstein et al., Phys.\ Rev.\ C 95 (2017) 034001.
%
\bibitem{HAU2014} F.~Hauenstein, PhD thesis, University of Erlangen-N\"{u}rnberg (2014)
%
\bibitem{JOW2016} S.~Jowzaee et al., Eur.\ Phys.\ J.\ A 52 (2016) 7.
%
\bibitem{FAL2017} G.~F\"{a}ldt and C.~Wilkin, Phys.\ Rev.\ C 95 (2017) 024004.
%

\end{thebibliography}
\end{document}